\documentclass[pra,aps,twocolumn,superscriptaddress,nofootinbib,showpacs]{revtex4}

\usepackage{graphicx}
\usepackage{amssymb}
\usepackage{amsmath}

\begin{document}

\title{More Really is Different}

\author{Mile Gu}
\address {Department of Physics, University of Queensland, St
Lucia, Queensland 4072, Australia.}

\author{Christian Weedbrook}
\address {Department of Physics, University of Queensland, St
Lucia, Queensland 4072, Australia.}

\author{\'{A}lvaro Perales}
\address {Department of Physics, University of Queensland, St
Lucia, Queensland 4072, Australia.}
\address {Dpto.\ Autom\'atica, Escuela Polit\'ecnica, Universidad de Alcal\'a, Alcal\'a de Henares, Madrid 28871, Spain.}

\author{Michael A. Nielsen}
\address {Department of Physics, University of Queensland, St
Lucia, Queensland 4072, Australia.}
\address{Perimeter Institute for Theoretical Physics, Waterloo, Ontario N2L 2Y5, Canada.}

\date{\today}

\begin{abstract}
  In 1972, P.~W.~Anderson suggested that `More is Different', meaning
  that complex physical systems may exhibit behavior that cannot be
  understood only in terms of the laws governing their microscopic
  constituents.  We strengthen this claim by proving that many
  macroscopic observable properties of a simple class of physical
  systems (the infinite periodic Ising lattice) cannot in general be
  derived from a microscopic description.  This provides evidence that
  emergent behavior occurs in such systems, and indicates that even if
  a `theory of everything' governing all microscopic interactions were
  discovered, the understanding of macroscopic order is likely to
  require additional insights.
\end{abstract}

\pacs{89.75.-k, 75.10.Hk}

\maketitle

\section{Introduction}

The reduction of collective systems to their constituent parts is
indispensable to science. The behavior of ideal gases can be
understood in terms of a simple model of non-interacting point
particles; the properties of chemical compounds predicted through
their underlying atomic structure; and much of the recent advances
in biology has been achieved by reducing biological behavior to
properties of the DNA molecule.

These and other triumphs have fostered the optimistic belief that all
scientific theories can ultimately be reduced to a small set of
fundamental laws; that the universe is broken up into a series of
reductive levels (e.g., ecosystems, multicellular living organisms,
cells, molecules, atoms, elementary particles); and that any
scientific theory that governs one reductive level can be
mathematically deduced from the laws that govern the reductive levels
below it~\cite{Weinberg93a,Anderson72a}. This encourages certain subfields
to claim a kind of moral high ground, based on an ideal of science as
determining the fundamental microscopic behavior, with the rest `just'
details.

Of course, many disagree that the rest is just details. In
1972, P.~W.~Anderson laid out such a case in his article ``More is
Different''~\cite{Anderson72a}, arguing that complex systems may
possess emergent properties difficult or impossible to deduce from a
microscopic picture. Anderson gives several examples which he suggests
illustrate this idea, based on broken symmetry, and goes so far as to
claim that in the limit of infinite systems, emergent principles take
over and govern the behavior of the system, which can no longer be
deduced from the behavior of the constituent parts. Since macroscopic
laws that govern macroscopic observables often implicitly assume this
infinite limit, they cannot logically be derived, even in principle,
from microscopic principles. Is Anderson correct? His examples were
largely speculative. The question of whether some macroscopic laws may
be fundamental statements about nature or may be deduced from some
`theory of everything' remains a topic of debate among
scientists~\cite{Laughlin00a,Weinberg93a}.

In this article we strengthen Anderson's claims by proving that
standard notions of reductionism cannot generally hold in a widely
studied class of collective systems, the infinite square Ising
lattice.  We show that for a large class of macroscopic observables,
including many of physical interest, the value of those observables is
formally undecidable, i.e., cannot generally be computed from the
fundamental interactions in the lattice.  Consequently, any
macroscopic law that governs the behavior of such properties cannot be
deduced from first principles.  Our result therefore indicates that perhaps a
`theory of everything' may not explain all natural phenomena;
additional experiments and intuition may be required at each reductive
level.

Our paper is inspired by previous results \cite{Wolfram85a,Moore90a,Schlijper97} on undecidability in physical systems. We employ a similar strategy, which is to map computational models into equivalent physical systems; the undecidability of the computational models then implies that there
must exist undecidable properties of those physical systems. Our proof
extends this mapping so that these undecidable properties encompass a large class of observables that are physically interesting on macroscopic scales. These results present analytical evidence for
emergence.

\section{Reductionism and the Periodic Ising Lattice}

Square Ising lattices describe a classical system of spins arranged at
the vertices of a $d$-dimensional rectangular grid. The state of each
spin is described by a single value ($0$ or $1$) and interacts only
with its $2d$ neighbors. In this paper, we work with planar lattices
($d = 2$), though our results easily generalize to higher dimensions.
While this simple model was first introduced to describe magnetic
materials~\cite{Chandler87a}, where each spin describes the
orientation of a microscopic magnetic moment, it has become ubiquitous
in modeling a diverse range of collective systems, including lattice
gases~\cite{Chandler87a}, neural activity~\cite{Rojas}, protein
folding~\cite{Bryngelson91}, and flocking~\cite{Toner95a}. Emergence
in such models would thus suggest it is of common occurrence in
nature. For convenience, we use the standard terminology of magnetism,
though our arguments apply equally to other applications of the model.

\begin{figure}[htp]
\centering
\includegraphics[width=0.50\textwidth]{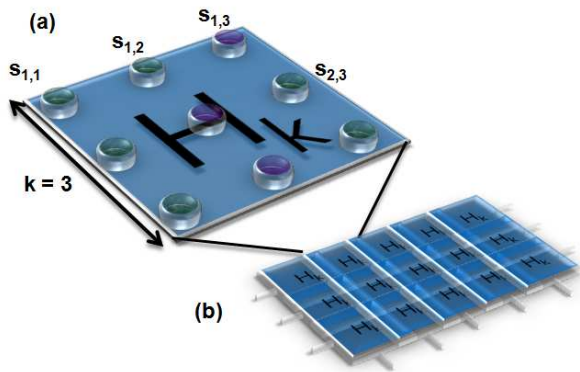}
\caption{\label{fig:thelattice} The square Ising lattice (a) consists of a
  rectangular grid of spins such that only adjacent spins interact,
  i.e., $c_{\mathbf{x},\mathbf{y}} = 0$ unless $|\mathbf{x} -
  \mathbf{y}| = \sum_i|x_i - y_i| = 1$.  Such a lattice is periodic if
  it can be specified completely by some Hamiltonian $H_k$ that acts
  on a $K \times K$ Ising block (b). Note that the Hamiltonians are
  tesselated in such a way that the adjacent blocks always share one
  common row or column.}
\end{figure}

Mathematically, we index each spin of the $2$-d square Ising lattice
by a vector of integers $\mathbf{x} = (i,j)$
(Fig.~\ref{fig:thelattice}(a)), such that $s_\mathbf{x} \in \{0,1\}$
denotes the state of the spin at location $\mathbf{x}$.  Interactions
on this lattice are described by the Hamiltonian $H$, a function that
maps each configuration of the lattice to a real number corresponding
to energy. The general Ising model with an external field has a
Hamiltonian of the form~\cite{Chandler87a}
\begin{align}
H = \sum c_{\mathbf{x},\mathbf{y}} s_\mathbf{x} s_\mathbf{y} + \sum M_\mathbf{x}
s_\mathbf{x}
\end{align}
where $c_{\mathbf{x},\mathbf{y}}$ are the interaction
energies between spins $s_\mathbf{x}$ and $s_\mathbf{y}$, and
$M_\mathbf{x}$ describes the external field at site $\mathbf{x}$. We
say spins $j$ and $k$ interact if $c_{j,k} \neq 0$. The ground states
of the system are configurations that minimize the value of $H$.

Consider a macroscopic system modeled by a square Ising lattice of $N
\times N$ spins, with $N \gg 1$.  Such systems often exhibit
periodicity, i.e., clusters of spins are often found to experience
similar interactions.  We can specify such systems by periodic Ising
models, which consist of a tessellation of spin blocks, each governed
by identical intra- and inter-block interactions
(Fig.~\ref{fig:thelattice}~(b)).

Understanding the behavior of such a macroscopic system need not
entail knowledge of the dynamics of each individual microscopic
constituent.  The physically relevant observables, at macroscopic
scales, such as magnetization (the proportion of spin in state $1$),
are generally global properties of the lattice. Insight into the
behavior of such systems may be obtained from knowledge of the
macroscopic laws that govern the dynamics of such properties.  While
\emph{a priori}, there is no guarantee that such laws should exist,
the existence of thermal physics and other macroscopic principles
suggests that the universe conspires in many instances to give the
macroscopic world some sort of order~\cite{Laughlin00a}.

In contrast, reductionism contends that any macroscopic order can be
understood by decomposing the system to its basic interactions, i.e.,
the known interactions of each periodic block within the lattice.
Thus, from a reductionist perspective, the fundamental science of such
a system is the determination of these interactions, and the rest is
just working out the consequences of those interactions.

We construct a class of periodic Ising models that directly contradict
this perspective. In particular, we consider $2$-d macroscopic
lattices where the spins of a $1$-d edge are fixed by some spatially
varying external magnetic field. We will show that at its lowest
energy state, a general class of macroscopic properties cannot be
generally predicted from knowledge of the lattice Hamiltonian $H_k$.
Thus any macroscopic law that governs these quantities must be
logically independent of the fundamental interactions.

In practice, of course, many periodic Ising systems are soluble. What relevance, then, do these results have for the practice of science? We
observe that in many cases of physical interest (e.g., the 3-d Ising
model), no explicit, formal solution is known; it is possible that
this is not merely a product of our ignorance, but rather because no
solution exists.

\section{The Approach}

Our approach is inspired by the existence of `emergent' phenomena in
mathematics. Unlike physical systems, the axioms that define a
mathematical system, its analogous `theory of everything', are
known; yet, many properties of such systems cannot be proven either
true or false, and hence are formally undecidable~\cite{godel31}. The
Turing machine~\cite{Turing36b} is one such system. First proposed to
formally describe a universal computer, Turing machines are
theoretical devices that consist of a finite state machine that
operates on an unbounded one-dimensional array of binary states.
Despite the fact that the behavior of these machines is formally
characterized, most questions regarding their long-term dynamics are
undecidable.

One well known example of undecidability is the halting
problem~\cite{Turing36b}, which asks whether a given machine ever
halts on a specific input. In fact, a much more general class of
questions is undecidable. Rice's theorem~\cite{Rice53a} states that
any non-trivial question about a Turing machine's black-box behaviour
is undecidable, i.e., any question about the functional relationship
between inputs and outputs.  For example, Rice's theorem
tells us that there is no general algorithm which will tell us whether
or not a given Turing machine acts to square its input, although of
course for specific machines it may be possible to determine whether
or not this is the case.

Numerous simple physical systems capable of simulating arbitrary
Turing machines have been proposed, e.g.,~\cite{Fred82,Moore90a}.
Since such `universal' systems are as powerful as Turing machines, and
thus an arbitrary computer, the only viable general method of
predicting the dynamics of such systems is by direct simulation. The only way to find whether or not it
halts is to run the machine \emph{ad infinitum}, there exists no
algorithm that can determine the eventual behavior of any universal
system.

The `Game of Life'~\cite{Conway82} is a well-known example. The state
of this system consists of an infinite $2$-dimensional rectangular
grid of cells, each of which is either alive or dead. The system
evolves in discrete time steps, where the fate of each cell depends on
the state of the eight cells in its neighborhood (i.e., the $3 \times
3$ block centered around the cell). Although this simple system
exhibits dynamics entirely defined only by a binary function (its update function) on nine
bits, it is universal. The `Game of Life' is not unique, and belongs
to a general class of discrete dynamical systems known as cellular
automata (CA), including Life without Death~\cite{Moore97b} and the
$1$-dimensional Rule 110~\cite{cook04a}.

The dynamics of a CA are governed by an update rule applied
identically to each cell, reminiscent of a periodic Ising lattice
where each block experiences the same Hamiltonian. This motivates
encoding the dynamics of a CA in the ground state of the periodic Ising lattice.
While such constructions exist \cite{Domany84a,barahona82a}, our constructions must
be tailored so computing the macroscopic properties of the lattices would
entail knowledge of the undecidable properties of the underlying CA.

\section{The Cellular Automata Encoding}

We encode the dynamics of any $d$-dimensional CA within the ground states of a $(d+1)$-dimensional periodic Ising lattice with a particular $H_K$ The construction is not unique; a given CA may be simulated by an infinite number of different periodic Ising lattices.

Formally, we consider a CA that consists of a $d$-dimensional lattice
of cells, each of which may be either $0$ or $1$. The neighborhood of
a cell is the set of cells in a block of cells $(2r+1)$ on a side, and
centered on the cell, where $r$ is some positive integer that
specifies the size of the neighborhood we are considering. The way the
state of a CA changes at each time-step is dictated by a local update
rule, i.e., a function, $f$, that maps this neighborhood to $\{0,1\}$.
For example, the state of any $1$-dimensional CA is defined by an
infinite array of binary numbers $\ldots b_{-1,t} b_{0,t} b_{1,t}
b_{2,t} \ldots$ at time $t$. If $r=1$, then at $t+1$, the state of
each cell updates according to $b_{k,t+1} =
f\left(b_{k-1,t},b_{k,t},b_{k+1,t}\right)$. In order to avoid
burdensome notation we will explicitly outline the mapping of a CA to
a periodic Ising lattice for the simple case of $d = r = 1$.  The
general mapping follows identical ideas.

\begin{figure}[htp]
\centering
\includegraphics[width=0.50\textwidth]{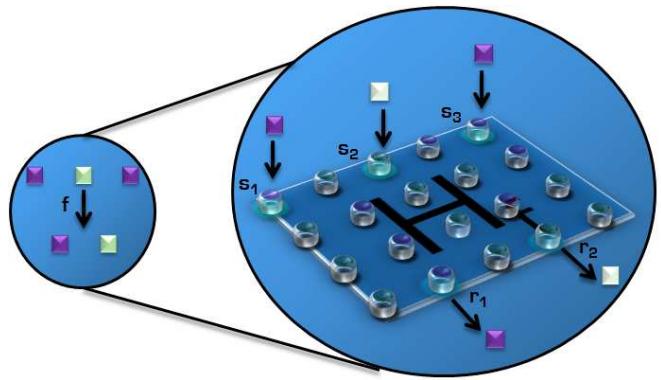}
\caption{\label{fig:designer} For any binary function $f$, we can construct
  an Ising block such that its ground state encodes $f$. If the input bits $s_i$ are fixed, then
  the output bits $r_i = f(s_i)$ when this block is at ground state.}
\end{figure}

We make use of `designer Ising blocks', bounded
$2$-dimensional blocks of spins with an associated Hamiltonian whose
ground state encodes a desired logical operation $f$.  The input is encoded in bits on one boundary of the block, while output bits on the boundary opposite (Fig.~\ref{fig:designer}). Formally, consider an
arbitrary binary function $f$ with $m$ inputs and $n$ outputs; we
define a `designer Ising block' as follows. Take a $C \times D$ block
of spins, where $C, D > \max(m,n)$, governed by a Hamiltonian $H_f$
with ground state set $\mathcal{G}_f$. We designate $m$ input spins,
$\overrightarrow{s} = (s_1, s_2, \ldots, s_m)$ from the first row to
encode the input and $n$ output spins, $\overrightarrow{r} = (r_1,r_2,
\ldots, r_n)$ from the last row as output.

We say a configuration of the lattice, $\mathbf{s}$, satisfies
$\{\overrightarrow{s},\overrightarrow{r}\}$ if the input and output
spins are in states $\overrightarrow{s}$ and $\overrightarrow{r}$
respectively. Suppose that (1) there exists $\mathbf{s} \in
\mathcal{G}_f$ that satisfies $\{\overrightarrow{s},\cdot\}$ for each
of the $2^m$ possible inputs of $f$ and (2) every $\mathbf{s} \in
\mathcal{G}_f$ satisfies $\{\overrightarrow{s},\overrightarrow{r} =
f(\overrightarrow{s})\}$, then we can set the ground state of the Ising
block to simulate the action of $f$ on any desired input by
appropriately biasing the input spins by external fields. In fact, previous results \cite{barahona82a} indicate appropriate blocks exist for any $f$; we outline an explicit method in the Appendix.

\begin{figure}[htp]
\centering
\includegraphics[width=0.50\textwidth]{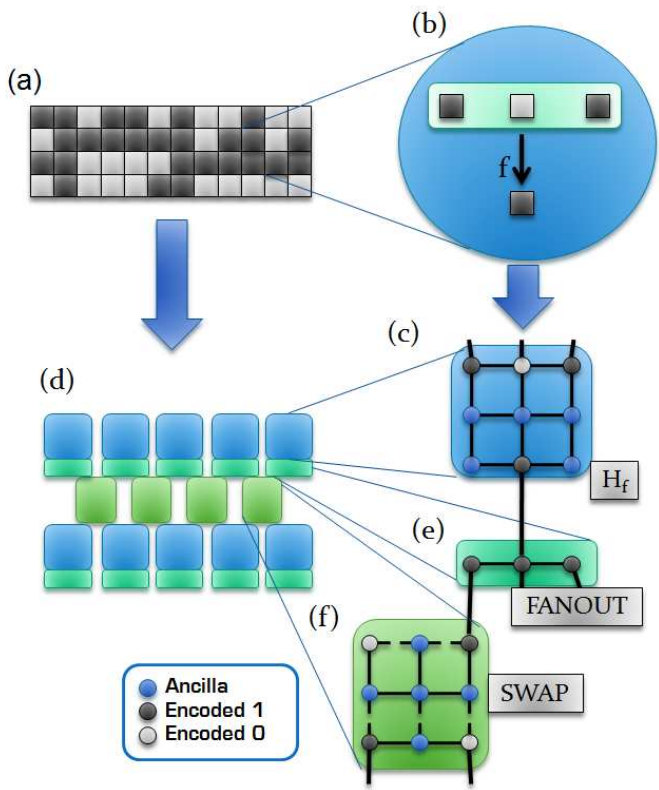}
\caption{\label{fig:cellmapping} (Color online) The dynamics of any given CA (a) with update
  rule $f$ (b) can be encoded in the ground state of a periodic Ising
  lattice (c) through a periodic tessellation of designer Ising blocks
  that simulate the operations $f$ (d), FANOUT (e), and SWAP (f).}
\end{figure}

To simulate the dynamics of a CA with an update function $f$, we
utilize designer Ising blocks that simulate (1) the update function
$f$; (2) the three way FANOUT function that takes a bit as input and
makes two copies; (3) the SWAP function, which switches the states of
its two inputs. Like the construction of a digital circuit these
building blocks can be tesselated together to simulate the dynamics of
any given CA (See Figure~\ref{fig:cellmapping}). The set of ground states of the resulting periodic Hamiltonian encodes
the dynamics of the given CA for all possible initial conditions. The
application of an external field to the first row (layer) of the
lattice then simulates the evolution of the encoded CA with a
particular initial condition. Thus, the ground state of the periodic
Ising model is universal.

\section{Undecidable Macroscopic Observables}

For each Turing machine, $T$, with input $x$, we can construct a
periodic Ising lattice such that knowledge of its ground state implies
complete knowledge of $T(x)$. Thus, the ground state of such lattices
must be undecidable. This result can be strengthened. Rice's theorem not only predicts that $T(x)$ is non-computable, but also all black box properties of $T$. Examples include, `is $T(x) > 500$? for all inputs' and `does $T$ double all its inputs'. Properties such as these must correspond to a property of the Ising lattice.

This motivates a \emph{Rice's theorem for physical systems}. Any property
of a physical system is specified by a function $P$ that maps the configuration space
of the system to the real numbers. Suppose the system is universal, and thus encodes an
underlying Turing machine $T$. Provided the observable property is
dependent on the output of $T(x)$, so that knowledge of $P$ implies
non-zero information about $T(x)$, then $P$ cannot be computable for
all such systems. This result is quite general. Given an Ising
lattice, there are infinitely many ways of encoding a Turing machine.
Provided a single one of these encodings affect the value of $P$, then
`Rice's theorem for physical systems' applies.

A useful example is the `prosperity' of a CA, the probability that a
randomly chosen cell at a random time step is alive. This
equates to the proportion of living cells, averaged over all time steps
from $0$ to infinity. In many universal CAs (Game of
Life, Life without Death), information is encoded in the presence or
absence of clusters of living cells of specific configurations,
referred to as gliders or ladders. Different computational results
lead to different numbers of gliders, and these gliders may cause
unbounded growth of living cells. Thus, the prosperity of a CA is
indeed dependent on the output of an encoded Turing machine, and must
be undecidable.

The prosperity of a CA is essentially a macroscopic observable --- for
a magnetic system, it is just the average magnetization of the system,
up to an additive constant.  Such observables are \emph{averaging
  properties}. That is, we can divide the Ising lattice into a
periodic tessellation of finitely sized blocks such that the property
depends on the average of some non-constant function $f$ on each
block. Formally, let $P: \mathcal{C} \rightarrow \mathbb{R}$ be a general
function that maps each configuration of the Ising lattice into a real
number, where $\mathcal{C}$ is the configuration space of the Ising
lattice. Divide the Ising lattice into a periodic tessellation of
finitely sized Ising blocks $B_1,B_2,\ldots$ of size $C \times D$, for
some fixed $C,D \in \mathbb{N}$. Let $\mathcal{C}_{C\times D}$ denote
the configuration space of each block. We introduce a non-trivial
function $f: \mathcal{C}_{C\times D} \rightarrow \mathbb{R}$, i.e.,
there exists $\mathbf{s}_1, \mathbf{s}_2 \in \mathcal{C}_{C\times D}$
such that $|f(\mathbf{s}_1) - f(\mathbf{s}_2)| \geq \epsilon$, for
some fixed $\epsilon > 0$. Define $\mathcal{A}(\mathbf{s}):
\mathcal{C} \rightarrow \mathbb{R}$, $\mathcal{A}(\mathbf{s}) =
\langle f(\mathbf{s}) \rangle$ as the average of $f$ over all $B_i$.

We say that $P$ is an averaging macroscopic property if knowledge of
$P(\mathbf{s})$ gives information about the value of
$\mathcal{A}(\mathbf{s})$ for some choice of $C$ and $D$. Explicitly,
let $\mathcal{R}_A$ be the range of $\mathcal{A}$ and $\mathcal{R}_P$
be the range of $P$. Suppose that for each $p \in \mathcal{R}_P$, $P(\mathbf{s}) = p$ implies $\mathcal{A}(\mathbf{s}) \not\in [a,b]$ for some non-zero interval $[a,b]$, then $P$ is an averaging macroscopic property. Total magnetization, average spin-spin correlation, and most standard quantities of physical interest can be
shown to fall into this category. Indeed, we will show that given such
a macroscopic property $P$, we construct a modified encoding scheme
such that the value of the given observable is almost entirely
dependent on the `prosperity' of the underlying CA.

The primary strategy is to replace the FANOUT blocks in our encoding
scheme with `magnifier blocks' (See Fig.~\ref{fig:magblock} a). The
`magnifier block' is a `designer Ising block' that simulates the
$3$-way FANOUT and additionally exhibits a ground state with notably
different contributions to $P$ depending on its input. Provided these
blocks are of sufficient size, knowledge of $P$ implies knowledge of
the average input of these magnifiers, i.e., the prosperity of the
underlying CA.

Formally, assume $P$ is decidable. In particular, the proposition `$P(\textbf{s}) = p$ at ground state $\textbf{s}$?" is decidable for any $p$. Then, there must exist an interval $[a,b]$ such that the proposition `$A(s)$ lies outside $[a,b]$ at ground state' is also decidable. However, since the Ising lattice is universal, a magnifier for any function exists. Therefore, we may construct a magnifier that ensures that $A(s) \in [a,b]$ iff the underlying prosperity is less than $1/2$. The decidability of $P$ then implies knowledge of the underlying prosperity. Hence, \emph{any such macroscopic property of the periodic Ising lattice is generally undecidable}. We illustrate this with a number of examples:

\begin{figure}[htp]
\centering
\includegraphics[width=0.50\textwidth]{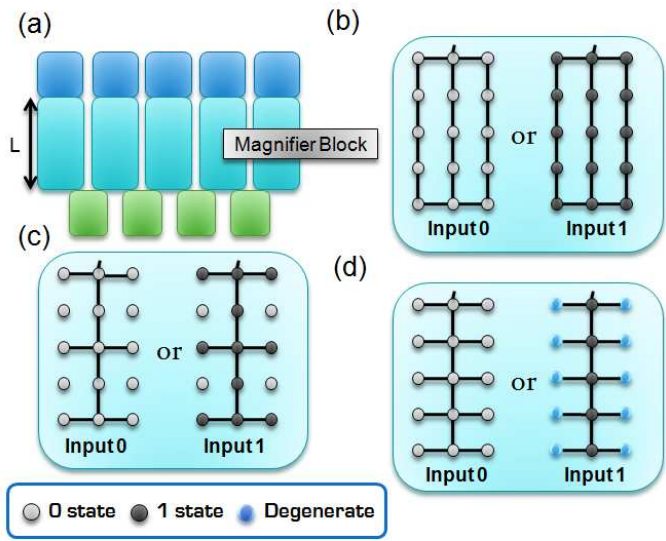}
\caption{\label{fig:magblock} Magnifier blocks can be inserted into the CA encoding (a) and made large enough such that their properties dominate the properties of the lattice. The ground state of these spin blocks (pictured) can exhibit significantly different values of magnetization (b), correlations functions (c) and degeneracy (d), for different inputs. Here, degenerate spins denote spins whose state has no effect on the energy of the lattice.}
\end{figure}

\begin{enumerate}

\item A magnetization magnifier has ground states of either all $0$'s
  or all $1$'s (Fig. \ref{fig:magblock} (b)). Thus, magnetization is
  undecidable.

\item The correlation length measures the scaling of $\lim_{r
    \rightarrow \infty} \langle s_{l,m},s_{l,m+r} \rangle$ (where
  $\langle \cdot \rangle$ denotes an average over all lattice sites)
  with $r$. Knowledge of the correlation length allows us to solve the
  undecidable question of whether the encoded CA will eventually have
  no living cells~\cite{Wolfram94a}. Thus the correlation length is
  undecidable.

\item Finite range correlations, i.e., $\langle s_{l,m},s_{l,m+r}
  \rangle$ or $\langle s_{l,m},s_{l+r,m} \rangle$, for some $r$,
  measure periodic structures. Since this property depends on the
  correlations of finitely sized blocks (magnified in Fig.~\ref{fig:magblock}~(c)), these correlations are undecidable.

\item The partition function at zero temperature is determined by the
  degeneracy of the system. Since degeneracy can be magnified,
  (Fig.~\ref{fig:magblock}~(d)) partition functions are
  non-computable.

\end{enumerate}

Chaitin~\cite{Chaitin82a} has emphasized that such undecidability
results automatically imply results about what is provable in such
systems.  In particular, our results imply that for any such
observable, there must \emph{exist} a specific Ising lattice for which
it is not possible to prove the ground state value of the observable.
The reason, in outline, is that if such a proof always existed, then
it would be possible to construct an algorithm for determining the
value of the observable, simply by enumerating and checking all
possible proofs.  We expect that this result readily generalizes to
lattices of finite temperatures and more exotic macroscopic
observables using different encodings and non-deterministic CAs.

\section{Discussion and Conclusion}

It may be objected that our results only hold in infinite lattices,
and hence are not relevant for real finite physical systems. Most
scientists would agree that any finite system, with finite energy,
exhibits behavior that is computable (but c.f.,~\cite{Penrose89a}).
Yet infinite systems also play an essential role in developing our
understanding of real physical systems. Even if we possessed a
supercomputer capable of simulating complex systems, we would still
not understand the system without referring to macroscopic concepts
such as phase transitions and the renormalization
group~\cite{Fisher98a}, which apply only in the limit of infinite
systems. Yet these same tools are essential to our understanding of the behavior of real physical systems.

In summary, Ising models play an important role in modeling many physical and biological
phenomena. Our results indicate that in such systems, many general
macroscopic ground state properties cannot be computed from
fundamental laws governing the microscopic constituents. Despite complete characterization of the system, we can assign two
different values to any such property, and there would exist no
logical way to prove which assignment is correct.  Instead, in
specific instances, the best one can do is assert the value of some
physically interesting properties as axiomatic, perhaps on the basis
of experimental evidence or (finite) simulations; this would truly be
an example where `more is different'.

Although macroscopic concepts are essential for understanding our
world, much of fundamental physics has been devoted to the search for
a `theory of everything', a set of equations that perfectly describe
the behaviour of all fundamental particles. The view that this is the
goal of science rests in part on the rationale that such a theory
would allow us to derive the behavior of all macroscopic concepts, at
least in principle. The evidence we've presented suggests this view
may be overly optimistic. A `theory of everything' is one of many
components necessary for complete understanding of the universe, but
is not necessarily the only one. The development of macroscopic laws from
first principles may involve more than just systematic logic, and could require
conjectures suggested by experiments, simulations or insight.

M.G., C.W. and M.A.N. acknowledge support from the Australian Research
Council. A.P. acknowledges financial support from Spanish MEC
(Programa para la Movilidad).  M.A.N. thanks Gerard Milburn and Cris
Moore for discussions. M.G. and C.W. thank Nick Menicucci, Niloufer
Johansen and Guifre Vidal for discussions.

\begin{figure}[htp]
\centering
\includegraphics[width=0.50\textwidth]{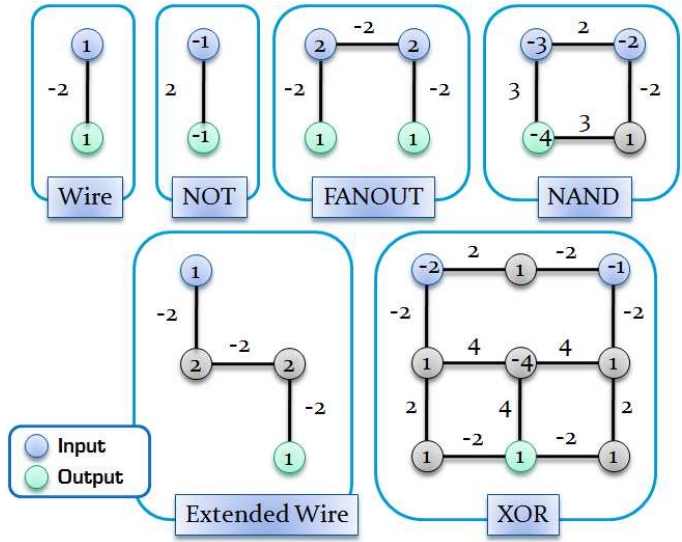}
\caption{\label{fig:appenfig1} (Color online) The interaction graphs of `designer
  Ising blocks' that simulate each of the basic boolean operations.
  These blocks can then be linked together by wires to simulate an
  arbitrary computation.}
\end{figure}

\begin{appendix}

\section{Universality of Ising Blocks}

In this section, we prove that the ground states of
designer Ising blocks are universal. Any boolean function $f$ can be represented by a logic circuit that consists of the following components: (1) wires (2) FANOUT gates
and (3) NAND gates. Mathematically, these operations are defined as
(1) $\mathrm{Wire}(b_1) = b_1$, (2) $\mathrm{FANOUT}(b_1) = (b_1,b_1)$
(3) $\mathrm{NAND}(b_1,b_2) = \neg(b_1 \wedge b_2)$.

We convert this to a \emph{planar circuit}, that is, one in which no
wires intersect. This is achieved by replacing each intersection
with a SWAP gate, $\mathrm{SWAP}(b_1,b_2) = (b_2,b_1)$. Such SWAP
gates can be decomposed into a network of three XOR gates i.e.,
$\mathrm{SWAP}(b_1,b_2) =
\mathrm{XOR}_1(\mathrm{XOR}_2(\mathrm{XOR}_1(b_1,b_2)))$, where
$\mathrm{XOR}_1(b_1,b_2) = (b_1 \oplus b_2, b_2)$ and
$\mathrm{XOR}_2(b_1,b_2) = (b_1 , b_1 \otimes b_2)$.

Observe that designer Ising blocks can be constructed to simulate of these components, i.e., (1) wires (2) FANOUT gates
(3) NAND gates and (4) XOR gates (c.f., Fig. \ref{fig:appenfig1}).
Therefore, any planar circuit, and hence any boolean function, can be
implemented by a designer Ising block.

\end{appendix}



\begin{thebibliography}{99}

\bibitem{Weinberg93a} Weinberg, S. \textit{Dreams of a Final Theory} (Pantheon
Books, 1993).

\bibitem{Anderson72a} Anderson, P.W. More Is Different. \textit{Science} \textbf{177,} 393 (1972).

\bibitem{Laughlin00a} Laughlin, R.B. and Pines, D. The Theory of Everything. \textit{Proc. Natl. Acad. Sci.
U.S.A.} \textbf{97,} 28 (2000).

\bibitem{Wolfram85a} Wolfram, S. Undecidability and intractability in theoretical physics. \textit{Phys. Rev. Lett.} \textbf{54,} 735 (1985).

\bibitem{Moore90a} Moore, C. Unpredictability and undecidability in dynamical systems. \textit{Phys. Rev. Lett.} \textbf{64,} 2354 (1990).

\bibitem{Schlijper97} Schlijper, A.G. Tiling problems and undecidability in the cluster variation method. \textit{J. Stat. Phys.} \textbf{50,} 689 (1988).

\bibitem{Chandler87a} Chandler, D. \textit{Introduction to modern statistical mechanics}
(Oxford University Press, New York, 1987).

\bibitem{Rojas} Rojas, R. \textit{Neural Networks - A Systematic Introduction}
(Springer, 1996).

\bibitem{Bryngelson91} Bryngelson, J.D. and Wolynes, P.G. Spin Glasses and the Statistical Mechanics of Protein Folding. \textit{Proceedings of the
National Academy of Sciences of the United States of
America} \textbf{84,} p. 7524 (1987).

\bibitem{Toner95a} Toner, J. and Tu, Y. Long-Range Order in a Two-Dimensional Dynamical XY Model: How Birds Fly Together. \textit{Phys. Rev. Lett.} \textbf{75,} 4326 (1995).

\bibitem{godel31} G${\rm \ddot{o}}$del, K. ${\rm \ddot{U}}$ber formal unentscheidbare Sätze der Principia Mathematica und verwandter Systeme I. \textit{Monatshefte f¨ur Math. u. Physik} \textbf{38,} 173
(1931).

\bibitem{Turing36b} Turing, A.M. On computable numbers, with an application to the Entscheidungsproblem. \textit{Proc. Lond. Math. Soc. 2} \textbf{42,} 230 (1936).

\bibitem{Rice53a} Rice, H.G. Classes of recursively enumerable sets and their decision problems. \textit{Trans. Amer. Math. Soc.} \textbf{74,} 358 (1953).

\bibitem{Fred82} Fredkin, E. and Toffoli, T. Conservative logic. \textit{International Journal of Theoretical
Physics} \textbf{21,} 219 (1982).

\bibitem{Conway82} Berlekamp, E. Conway, J. and Guy, R. \textit{Winning Ways
for your mathematical plays} (Academic Press, Oxford, 1982).

\bibitem{Moore97b} Griffeath, D. and Moore, C. Life without death is P-complete. \textit{Complex Systems} \textbf{10,} 437
(1996).

\bibitem{cook04a} Cook, M. Universality in Elementary Cellular Automata. \textit{Complex Systems} \textbf{15,} (2004).

\bibitem{Domany84a} Domany, E. and Kinzel, W. Equivalence of Cellular Automata to Ising Models and Directed Percolation. \textit{Phys. Rev. Lett.} \textbf{53,} 311
(1984).

\bibitem{barahona82a} Barahona, F. On the computational complexity of Ising spin glass models. \textit{J. Phys. A: Math. Gen.} \textbf{15,} 3241 (1982).

\bibitem{Wolfram94a} Wolfram, S. \textit{Cellular automata and complexity} (Addison-
Wesley, Redwood City, CA, 1994).

\bibitem{Chaitin82a} Chaitin, G.J. G${\rm \ddot{o}}$del's theorem and information. \textit{International Journal of Theoretical
Physics} \textbf{22,} 941 (1982).

 \bibitem{Penrose89a} Penrose, R. \textit{The emperor's new mind}. (Oxford
  University Press, Oxford, 1989).

 \bibitem{Fisher98a} Fisher, M.E. Renormalization group theory: Its basis and formulation in statistical physics. \textit{Rev. Mod. Phys.} \textbf{70,} 653 (1998).


\end{thebibliography}
\end{document}